\documentclass[10pt]{iopart}

\usepackage{amstext}
\usepackage{graphicx}

\begin{document}

\title{Correlation Analysis With Scale-local Entropy Measures}

\author{J G Reid and T A Trainor}

\address{CENPA, Box 354290, University of Washington, Seattle, Washington 98195-4290}

\ead{trainor@hausdorf.npl.washington.edu}

\begin{abstract}

A novel method for correlation analysis using scale-dependent R\'enyi entropies is described.  The method involves calculating the entropy of a data distribution as an explicit function of the scale of a $d$-dimensional partition of $d$-cubes, which is dithered to remove bias.  Analytic expressions for dithered scale-local entropy and dimension for a uniform random point set are derived and compared to Monte Carlo results.  Simulated nontrivial point-set correlations representing condensation and clustering are similarly analyzed.
\vskip .2in
keywords: scale, entropy, dimension, information, fractal, complex system, phase transition
\end{abstract}


\submitto{\JPA}

\maketitle

%:::::::::::::::::::::::::::::::::::

\section{Introduction}

Entropy is a measure of the {\em size} of a data distribution contained within a bounded region (distribution support) of some space.  In a thermodynamic context this distribution size is interpreted as the number of quantum states accessible to a dynamical system given macroscopic constraints.  More generally, if a measure space is partitioned, the measure distribution size is estimated by the effective number of partition elements given the distribution weighting.  

Definition of the space partition is a central element of entropy calculation.  The partition is sometimes defined as a small-scale limiting partition of the space ({\em e.g.,} thermodynamic limit, limit procedures in classical analysis \cite{KandK}), sometimes based on properties of the data distribution itself and/or on the analysis goals (as in wavelet analysis \cite{Wavelet}).  

We define entropy as an explicit function of the partition definition, a scaled binning of the measure space.  We calculate the entropy and related quantities as functions of the partition scale(s), similar to the multi-scale partition approaches used in fractal dimension calculations \cite{Grassberger} and image deconvolution \cite{Pantin}. Unlike some analysis methods we invoke no scale limits. Quantities are defined on bounded scale intervals explictly excluding asymptotic limits; the analysis system is in this sense {\em scale local}.

The end result of this approach is an entropy which represents {\em arbitrary} data correlations as a distribution on scale, particularly useful for problems where the detailed scaling behavior of correlations over substantial scale intervals is of interest ({\em e.g.,} condensation, coalescence, critical phenomena, strange attractors), where instrumental effects may distort scale distributions, and where the correlation structure is not simply expressible as a power law or other elementary function. \cite{Thesis}

In this paper we describe precision binning methods, define the basic scale-local entropy measure and generalize other aspects of information theory to define the scale-dependent entropy difference or information between an object distribution and a model reference as a differential correlation measure. Based on scale derivatives of entropy and information we define scale-local dimension and dimension transport as generalizations of conventional counterparts based on limit concepts. We apply these correlation measures to several simulations and real data analysis problems.

\section{Scale-local Entropy}

The entropy definition employed here is based on $C_q(e)$, the rank-$q$ correlation integral at scale $e$.  Given a data distribution in a $d$-dimensional primary measure space spanned by variables $\{x_i\}$, we consider a set of corresponding correlation spaces containing all possible $q$-point clusters of data points ($q$-tuples).  There is one such $q$-point distribution (in a $q$-fold Cartesian product space) for each unique $q$ value.  The $q$-point correlation integral is the projection of the $q$-point distribution onto its difference subspace spanned by $\{x_i-x_j\}$, integrating over the sum variable(s). \cite{SLTM}  The integration limit of the correlation integral on the difference variables is in the simplest case (isotropic binning) the single scale of the analysis. 

The reciprocal of the correlation integral estimates the {\em effective} bin number in the $d(q-1)$-dimensional difference subspace. It's counterpart in the primary measure space is the $(q-1)^{th}$ root, the effective bin number in the primary space.  Defining entropy as the logarithm of the effective primary-space bin number is consistent with entropy as a logarithmic size measure and is a generalization of the thermodynamic definition, the logarithm of the number of accessible states. The correlation integral can be approximated by binning the primary measure space, and expressed in terms of normalized bin contents $p_{i}(e)$, in which case $C_q(e) \simeq \sum_{i=1}^{M(e)} p_{i}(e)^q $.  This results in the rank-$q$ R\'enyi entropy \cite{Renyi}
%%%%%%%%%%%%%%%%%%%%%%%%%%%%%%%%%%%%%%%%%%%%%%%%%%%%%%%%%%%%%%%
\begin{equation}
S_q(e) \equiv \log \left[ {C_q(e)}^{-\frac{1}{q-1}}\right] \simeq \frac{1}{1-q}\log \left[\sum_{i=1}^{M(e)}p_{i}(e)^q\right]. 
\end{equation}
%%%%%%%%%%%%%%%%%%%%%%%%%%%%%%%%%%%%%%%%%%%%%%%%%%%%%%%%%%%%%%%
The entropy $S_q$, the number of occupied bins $M$, and the bin probability $p_i$ are explicit functions of the binning scale $e$. More generally, a non-isotropic binning (one utilizing bins without unit aspect ratio) would imply a multidimensional scale space.

\section{Scale-local Information}

Given scale-local entropy we define scale-local information as a basis for differential comparisons between object and reference distributions, data and model.  There are a number of possible information definitions from information theory and topology, with significant differences in performance.  We define information here as the difference between entropies for object and reference distributions. This implies that the effective bin number for an object distribution is compared {\em in ratio} to that of a reference. Nonzero information implies {\em multiplicative} reduction of effective bin number by increased correlation structure in the object distribution relative to the reference (cumulant analysis is an alternative differential approach emphasizing {\em linear or additive} reduction of distribution size). Scale-local information is then defined as
%%%%%%%%%%%%%%%%%%%%%%%%%%%%%%%%%%%%%%%%%%%%%%%%%%%%%%%%%%%%%%%
\begin{eqnarray}
I_q(e)&=&S_{q, \text{ref}}(e) - S_{q, \text{obj}}(e).
\end{eqnarray}
%%%%%%%%%%%%%%%%%%%%%%%%%%%%%%%%%%%%%%%%%%%%%%%%%%%%%%%%%%%%%%%
Information so defined provides a differential comparison between a data or {\em object} distribution and a model or {\em reference}.  For example, the reference distribution for an arbitrary point set would be a distribution with the same number of points which maximally `fills' the bounded support -- a uniform random distribution.  Since the uniform reference is useful in many applications we derive explicit analytic forms for its entropy and dimension.

\section{Binning and Dithering}

Although these scale-local analysis methods are completely general as to the nature of the measure distribution, for the purpose of illustration we emphasize point sets in this paper. The measure distribution is then a set of points in a $d$-dimensional embedding space ({\em e.g.,} the distribution of particles from a heavy-ion collision in momentum space).  The analysis begins by applying a partition to the embedding space.  For algebraic simplicity we consider a grid of $d$-dimensional cubes, an isotropic binning.  At each scale there is a continuum of possibilities for the relative position of the binning system on the embedding space.  Differing partition placement effects the analysis in general, and there is no {\em a priori} reason to prefer any single placement.  Thus, we average (dither) over all partition placements to calculate the entropy at each scale.

We define a dithering phase $\phi$ for each of the $d$ embedding-space dimensions.  The relationship between the partition system and the measure distribution is controlled by varying $\phi$.  To implement dithering we calculate the correlation integral $\sum_{i=1}^{M(e)}p_{i}(e)^q$ of an event $J$ times at each scale, incrementing $\phi$ each time.  Finally we average over these results to obtain the entropy.  The dithered entropy $S_q(e)$ is thus
%%%%%%%%%%%%%%%%%%%%%%%%%%%%%%%%%%%%%%%%%%%%%%%%%%%%%%%%%%%%%%%
\begin{eqnarray}
S_q(e)&=&\frac{1}{1-q}\log \left[\left\langle\sum_{i=1}^{M_{\phi}(e)}p_{i,\phi}^q(e)\right\rangle_{\phi}\right]. 
\end{eqnarray}
%%%%%%%%%%%%%%%%%%%%%%%%%%%%%%%%%%%%%%%%%%%%%%%%%%%%%%%%%%%%%%%
where $p_{i,\phi}(e)$ is the bin probability of the $i$th bin for binning phase $\phi$ and scale $e$.  $p_i$ is normalized so that $\sum_i p_i=1$ at each scale $e$ and dithering phase $\phi$, the sum taken over all occupied bins in each partition.

\section{A Simple Example}

\begin{figure}[ht]
%\centereps{4in}{2.16in}{2d50k.eps}
\centering
\includegraphics[width=4in]{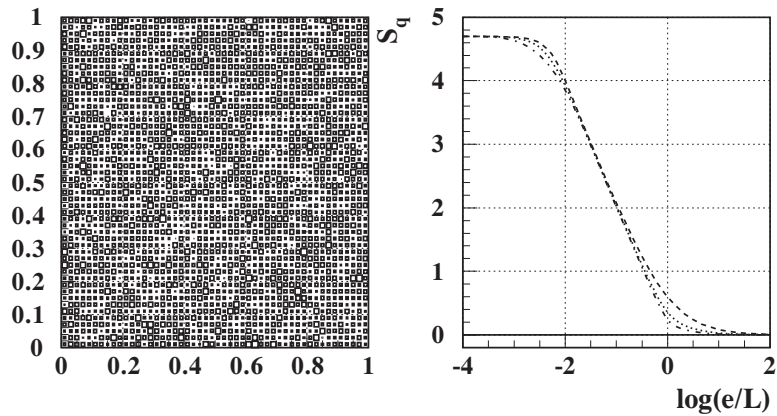}
\caption{Results of scale-local entropy analysis applied to a distribution of 50,000 random points uniformly distributed on a 2D unit square with $L=1$.  A box plot of the distribution itself is shown in the left panel, the right panel shows the measured entropy for $q=0$ (dashed line), $q=2$ (dotted line), and $q=5$ (dot-dashed line).}
\label{1dRGUD}
\end{figure}

A simple application of scale-local entropy and dithered binning illustrates the analysis process.  We obtain the scaled entropy of a 2D uniform distribution of $N$ randomly generated points on a unit-square support for several values of index $q$. A Monte Carlo simulation with 50k points is shown in figure~\ref{1dRGUD} along with analysis results for $q=0,2,5$.
%%%%%%%%%%%%%%%%%%%%%%%%%%%%%%%%%%%%%%%%%%%%%%%%%%%%%%%%%%%
%  1
% analysis of 2d uniform random distribution (50k pts)
% 
%
%%%%%%%%%%%%%%%%%%%%%%%%%%%%%%%%%%%%%%%%%%%%%%%%%%%%%%%%%%%
To interpret these results we consider small-, intermediate-, and large-scale regimes.  In the small-scale limit, well below the point-separation scale ($e \ll L/\sqrt{N}$), each distribution point occupies a single bin; there are N occupied bins ($M=N$), each with bin probability $p_i=1/N$.  Thus, at small scale the entropy approaches
%%%%%%%%%%%%%%%%%%%%%%%%%%%%%%%%%%%%%%%%%%%%%%%%%%%%%%%%%%%%%%%
\begin{equation}
\lim_{e\rightarrow 0}S_q(e)=\frac{1}{1-q}\log \left[\sum_{i=1}^{N}\left(\frac{1}{N}\right)^q\right]=\log N. 
\end{equation}
%%%%%%%%%%%%%%%%%%%%%%%%%%%%%%%%%%%%%%%%%%%%%%%%%%%%%%%%%%%%%%%
At intermediate scales we idealize the uniform point set to a continuum ($N\rightarrow \infty$). The number of occupied bins is then simply $M = (L/e)^2$ and the bin probability is $p_i = 1/M$ giving $S_q = 2\, \log(L/e)$. At scales substantially greater than the boundary scale ($e \gg L$) the entire distribution is contained in a single bin: $M=1$ and $p_i=1$.  The rank-$q$ entropy thus vanishes at scales much larger than the distribution boundary size. The detailed $q$-dependence near the particle-separation and boundary scales is amenable to an analytic treatment.

\section{Derivation of $S_q(e)$ for a Bounded Uniform Distribution}

Information is a relative quantity, a matter of definition.  It is impossible to make an absolute determination of the information content of an arbitrary distribution.  Using a maximum-entropy reference we can measure information relative to a distribution which is minimally correlated (given certain constraints).  This motivates us to derive an algebraic form for the scale-local entropy of a {\em bounded uniform distribution} (BUD).  This distribution represents a maximum-entropy hypothesis within a boundary, a maximum filling of the distribution support.  This is a correlation reference from which any object distribution may deviate with reduced entropy.\footnote{Boundedness is itself a form of correlation in a self-consistent description.} The derivation is presented in two parts: scales below and above the boundary scale.

\subsection{Below the boundary scale: $e \le L$}

For partitions below the boundary scale there is at least one bin in the interior of the embedding space.  To derive an analytic form for the entropy of a BUD we consider a two-dimensional distribution (and later generalize to $d$ dimensions).  The BUD is defined on a square support with side length $L$.  Because the distribution is uniform the probability of finding a point in any given bin is simply determined by the bin area. 
%%%%%%%%%%%%%%%%%%%%%%%%%%%%%%%%%%%%%%%%%%%%%%%%%%%%%%%%%%%
%  2
% Dither Variables 2d
% 
\begin{figure}[ht]
%\centereps{4in}{4in}{DitherVariables2d.eps}
\centering
\includegraphics[width=4in]{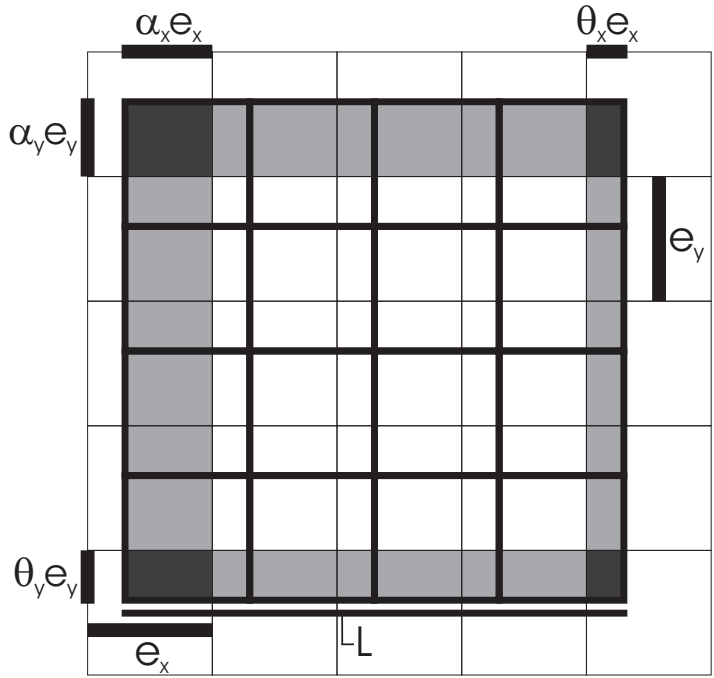}
\caption{A binning cartoon that illustrates the relationships among the binning, distribution support and dithering variables for the case of a two-dimensional embedding space.}
\label{DitherVariables2d}
\end{figure}
%
%%%%%%%%%%%%%%%%%%%%%%%%%%%%%%%%%%%%%%%%%%%%%%%%%%%%%%%%%%%
Figure~\ref{DitherVariables2d} shows a BUD binned with a general rectangular binning (thin dark lines).  We calculate the bin contents for each bin (as a function of scale) and integrate over all possible dithering configurations to determine the analytic form of the entropy.  For bins that are entirely contained within the embedding space (interior bins) the bin probability is trivial: $p_i = e_{\text{x}}e_{\text{y}}/L^2$ is the area of the bin divided by the total area of the support, independent of bin dithering.  For edge and corner bins the problem is more complicated. We consider each bin type -- interior (white), edge (light grey) and corner (dark grey) -- separately.  

Corner bins contain both an x- and y-axis support edge.  For all dithering configurations there are four corner bins.  The effective area of these corner bins is $e_{\text{x}} e_{\text{y}}$ scaled down by the fraction of the bin along the $i$th axis ($i \in [1,d]$) that overlaps the data.  By definition, the phase along an axis is $\phi_i=(1-\alpha_i)$, where $\alpha_i$ is the bin overlap fraction for the first bin on axis $i$ ($\phi \in [0,1]$).  We define $\Delta_i\equiv (L/e_i)-\text{int} (L/e_i)$ and express the amount of overlap between the last bin on axis $i$ and the edge of the support along that axis as $\theta_i=\phi_i+\Delta_i-\text{int} (\phi_i+\Delta_i)$.  With these definitions, calculating the contribution to the correlation integral of the corner bins is a matter of integrating over all $\phi$ values using the relevant bin probabilities.  Labeling the corner bins from right to left starting with the upper left bin we write down the $\phi$-dependent corner bin probabilities as
%%%%%%%%%%%%%%%%%%%%%%%%%%%%%%%%%%%%%%%%%%%%%%%%%%%%%%%%%%%%%%%
\begin{eqnarray}
p_{\text{A}}^q&=&(1-\phi_{\text{x}})^q(1-\phi_{\text{y}})^q\left( \frac{e_{\text{x}}e_{\text{y}}}{L^2}\right)^q \\ \nonumber
p_{\text{B}}^q&=&(1-\phi_{\text{x}})^q[\phi_{\text{y}}+\Delta_{\text{y}}-\text{int} (\phi_{\text{y}}+\Delta_{\text{y}})]^q\left( \frac{e_{\text{x}}e_{\text{y}}}{L^2}\right)^q \\ \nonumber
p_{\text{C}}^q&=&[\phi_{\text{x}}+\Delta_{\text{x}}-\text{int} (\phi_{\text{x}}+\Delta_{\text{x}})]^q(1-\phi_{\text{y}})^q\left( \frac{e_{\text{x}}e_{\text{y}}}{L^2}\right)^q \\ \nonumber
p_{\text{D}}^q&=&[\phi_{\text{x}}+\Delta_{\text{x}}-\text{int} (\phi_{\text{x}}+\Delta_{\text{x}})]^q[\phi_{\text{y}}+\Delta_{\text{y}}-\text{int} (\phi_{\text{y}}+\Delta_{\text{y}})]^q\left( \frac{e_{\text{x}}e_{\text{y}}}{L^2}\right)^q.
\end{eqnarray}
%%%%%%%%%%%%%%%%%%%%%%%%%%%%%%%%%%%%%%%%%%%%%%%%%%%%%%%%%%%%%%%

We calculate the dither-averaged correlation integral $C_q=<\sum_i p_i^q>_{\phi}$ by integrating over the different dithering configurations for each bin and summing $<\sum_i p_i^q>_{\phi}=\sum_i <p_i^q>_{\phi}$.  Following this approach we integrate the $p_i^q$ expressions over the two dithering variables and sum results to calculate the correlation integral.  The $\phi$ integral over the first corner bin yields
%%%%%%%%%%%%%%%%%%%%%%%%%%%%%%%%%%%%%%%%%%%%%%%%%%%%%%%%%%%%%%%
\begin{eqnarray}
<p_{\text{A}}^q>_{\phi}&=&\left( \frac{e_{\text{x}}e_{\text{y}}}{L^2}\right)^q \int_0^1\int_0^1(1-\phi_{\text{x}})^q(1-\phi_{\text{y}})^q\, \rmd\phi_{\text{x}}\, \rmd\phi_{\text{y}} \\ \nonumber
&=&\left(\frac{e_{\text{x}}e_{\text{y}}}{L^2}\right)^q\left(\frac{1}{1+q}\right)^2.
\end{eqnarray}
%%%%%%%%%%%%%%%%%%%%%%%%%%%%%%%%%%%%%%%%%%%%%%%%%%%%%%%%%%%%%%%
The second term is
%%%%%%%%%%%%%%%%%%%%%%%%%%%%%%%%%%%%%%%%%%%%%%%%%%%%%%%%%%%%%%%
\begin{eqnarray}
<p_{\text{B}}^q>_{\phi}&=&\left(\frac{e_{\text{x}}e_{\text{y}}}{L^2}\right)^q \int_0^1\int_0^1 (1-\phi_{\text{x}})^q[\phi_{\text{y}}+\Delta_{\text{y}}-\text{int}(\phi_{\text{y}}+\Delta_{\text{y}})]^q\, \rmd\phi_{\text{x}}\, \rmd\phi_{\text{y}}  \\ \nonumber
&=&\left(\frac{e_{\text{x}}e_{\text{y}}}{L^2}\right)^q\left(\frac{1}{1+q}\right)\left[\int_{\Delta_{\text{y}}}^1 v^q\, \rmd v+\int_0^{\Delta_{\text{y}}} v^q\, \rmd v\right] \\ \nonumber
&=&\left(\frac{e_{\text{x}}e_{\text{y}}}{L^2}\right)^q\left(\frac{1}{1+q}\right)^2.
\end{eqnarray}
%%%%%%%%%%%%%%%%%%%%%%%%%%%%%%%%%%%%%%%%%%%%%%%%%%%%%%%%%%%%%%%
The third and fourth terms are similar to the first and second; each of the four corner bins contributes an $(e_{\text{x}}e_{\text{y}}L^{-2})^q(1+q)^{-2}$ term to the total dither-averaged correlation integral.  

The contributions from edge bins are simpler to calculate.  The overlap fraction is unity in the direction parallel to the support edge; along that axis we merely count the number of edge bins.  The integral along the second axis is similar to the corner bins.  Again there are four terms, but symmetry simplifies the problem
%%%%%%%%%%%%%%%%%%%%%%%%%%%%%%%%%%%%%%%%%%%%%%%%%%%%%%%%%%%%%%%
\begin{eqnarray}
<p_{\text{edge}}^q>_{\phi}&=&\left(\frac{e_{\text{x}}e_{\text{y}}}{L^2}\right)^q \int_0^1\int_0^1 \left[\text{int}\left(\frac{L}{e_{\text{x}}}+1+\phi_{\text{x}}\right)-2\right] (1-\phi_{\text{y}})^q\, \rmd\phi_{\text{x}}\, \rmd\phi_{\text{y}} \\ \nonumber
&=&\left(\frac{e_{\text{x}}e_{\text{y}}}{L^2}\right)^q \int_0^1\int_0^1 \left[\text{int}\left(\frac{L}{e_{\text{x}}}+1+\phi_{\text{x}}\right) -2\right] [\phi_{\text{y}}+\Delta_{\text{y}}-\text{int}(\phi_{\text{y}}+\Delta_{\text{y}})]^q \, \rmd\phi_{\text{x}}\, \rmd\phi_{\text{y}} \\ \nonumber
&=&\left(\frac{e_{\text{x}}e_{\text{y}}}{L^2}\right)^q\left(\frac{L}{e_{\text{x}}}-1\right)\left(\frac{1}{1+q}\right).
\end{eqnarray}
%%%%%%%%%%%%%%%%%%%%%%%%%%%%%%%%%%%%%%%%%%%%%%%%%%%%%%%%%%%%%%%
which is the expression for the x-axis border bins. Contributions from the y-axis bins are obtained by switching indices.  The contribution from edge bins is thus
%%%%%%%%%%%%%%%%%%%%%%%%%%%%%%%%%%%%%%%%%%%%%%%%%%%%%%%%%%%%%%%
\begin{equation}
2\left(\frac{e_{\text{x}}e_{\text{y}}}{L^2}\right)^q\left[\left(\frac{L}{e_{\text{x}}}-1\right)\left(\frac{1}{1+q}\right)+\left(\frac{L}{e_{\text{y}}}-1\right)\left(\frac{1}{1+q}\right)\right].
\end{equation}
%%%%%%%%%%%%%%%%%%%%%%%%%%%%%%%%%%%%%%%%%%%%%%%%%%%%%%%%%%%%%%%
There remains the integral over interior bins, a simple matter of bin counting
%%%%%%%%%%%%%%%%%%%%%%%%%%%%%%%%%%%%%%%%%%%%%%%%%%%%%%%%%%%%%%%
\begin{equation}
<p_{\text{interior}}^q>_{\phi} = \left(\frac{e_{\text{x}}e_{\text{y}}}{L^2}\right)^q\left(\frac{L}{e_{\text{x}}}-1\right)\left(\frac{L}{e_{\text{y}}}-1\right).
\end{equation}
%%%%%%%%%%%%%%%%%%%%%%%%%%%%%%%%%%%%%%%%%%%%%%%%%%%%%%%%%%%%%%%

The full correlation integral can be assembled from the corner, edge, and interior bin integrals
%%%%%%%%%%%%%%%%%%%%%%%%%%%%%%%%%%%%%%%%%%%%%%%%%%%%%%%%%%%%%%%
\begin{eqnarray}
\fl C_q(e)= \; <\sum_i p_i^q>_{\phi}=\sum_i <p_i^q>_{\phi} \\ \nonumber
\fl \;\;\;\;\;\;\;\;\;\, =\left(\frac{e_{\text{x}}e_{\text{y}}}{L^2}\right)^q \left[4\left(\frac{1}{1+q}\right)^2+2\left(\frac{1}{1+q}\right)\left(\frac{L}{e_{\text{x}}}-1\right)+2\left(\frac{1}{1+q}\right)\left(\frac{L}{e_{\text{y}}}-1\right)+\left(\frac{L}{e_{\text{x}}}-1\right)\left(\frac{L}{e_{\text{y}}}-1\right)\right] \\ \nonumber
\fl \;\;\;\;\;\;\;\;\;\, =\left(\frac{e_{\text{x}}e_{\text{y}}}{L^2}\right)^{q-1} \left[1+\left(\frac{1-q}{1+q}\right)\left(\frac{e_{\text{x}}}{L}\right)\right]\left[1+\left(\frac{1-q}{1+q}\right)\left(\frac{e_{\text{y}}}{L}\right)\right].
\end{eqnarray}
%%%%%%%%%%%%%%%%%%%%%%%%%%%%%%%%%%%%%%%%%%%%%%%%%%%%%%%%%%%%%%%
Inserting this result into the definition of the rank-$q$ entropy we find that
%%%%%%%%%%%%%%%%%%%%%%%%%%%%%%%%%%%%%%%%%%%%%%%%%%%%%%%%%%%%%%%
\begin{eqnarray} \label{genent}
\fl S_q(e)=\frac{1}{1-q} \log [C_q(e)] \\ \nonumber
\fl \;\;\;\;\;\;\;\;\;\, = \frac{q-1}{1-q} \log \left(\frac{e_{\text{x}}e_{\text{y}}}{L^2}\right)+\frac{1}{1-q} \log \left[1+\left(\frac{1-q}{1+q}\right)\left(\frac{e_{\text{x}}}{L}\right)\right]+\frac{1}{1-q} \log \left[1+\left(\frac{1-q}{1+q}\right)\left(\frac{e_{\text{y}}}{L}\right)\right] \\ \nonumber
\fl \;\;\;\;\;\;\;\;\;\, = \left\{\log\left(\frac{L}{e_{\text{x}}}\right)+\frac{1}{1-q}\log\left[1+\left(\frac{1-q}{1+q}\right)\frac{e_{\text{x}}}{L}\right]\right\}+\left\{\log\left(\frac{L}{e_{\text{y}}}\right)+\frac{1}{1-q}\log\left[1+\left(\frac{1-q}{1+q}\right)\frac{e_{\text{y}}}{L}\right]\right\}.
\end{eqnarray}
%%%%%%%%%%%%%%%%%%%%%%%%%%%%%%%%%%%%%%%%%%%%%%%%%%%%%%%%%%%%%%%
Because of the additivity of entropy under product this entropy expression contains two terms, one for each axis, which suggests the $d$-dimensional generalization: adding an equivalent term for each additional dimension.

\subsection{Above the boundary scale: $e \ge L$} 

For the derivation of Eq. (\ref{genent}) we considered contributions from three types of bins: corner, border, and interior.  In the 2D derivation this approach is only valid when both $e_{\text{x}} \le L$ and $e_{\text{y}} \le L$.  If either of the scales is larger than the support size there are no {\em interior} bins and Eq. (\ref{genent}) is not valid.  Thus, to obtain the entropy expression for a BUD at large scale consider a single axis (we now exploit the additivity of entropy with respect to dimension for uncorrelated systems) with $e \ge L$.  When there is a bin edge in the support there are exactly two corner bins; we express the bin probability of the second bin in terms of the first
%%%%%%%%%%%%%%%%%%%%%%%%%%%%%%%%%%%%%%%%%%%%%%%%%%%%%%%%%%%%%%%
\begin{eqnarray}
p_1^q&=&\left[\alpha\left(\frac{e}{L}\right)\right]^q \\ \nonumber
p_2^q&=&[1-p_1]^q.
\end{eqnarray}
%%%%%%%%%%%%%%%%%%%%%%%%%%%%%%%%%%%%%%%%%%%%%%%%%%%%%%%%%%%%%%%%
When the support fits completely inside a single bin $p_0=1$; the distance from the support edge to the nearest bin edge is larger than the size of the support ($\alpha e \ge L$). We now evaluate the relevant integrals
%%%%%%%%%%%%%%%%%%%%%%%%%%%%%%%%%%%%%%%%%%%%%%%%%%%%%%%%%%%%%%%
\begin{equation}
<p_{0}^q>_{\alpha}=\int_{\frac{L}{e}}^1 [1]^q\, \rmd\alpha=1-\frac{L}{e},
\end{equation}
%%%%%%%%%%%%%%%%%%%%%%%%%%%%%%%%%%%%%%%%%%%%%%%%%%%%%%%%%%%%%%%  
%%%%%%%%%%%%%%%%%%%%%%%%%%%%%%%%%%%%%%%%%%%%%%%%%%%%%%%%%%%%%%%
\begin{equation}
<p_{1}^q>_{\alpha}=\int_0^{\frac{L}{e}} \alpha^q\left(\frac{e}{L}\right)^q\, \rmd\alpha =\frac{L}{e}\left(\frac{1}{1+q}\right)=<p_{2}^q>_{\alpha},
\end{equation}
%%%%%%%%%%%%%%%%%%%%%%%%%%%%%%%%%%%%%%%%%%%%%%%%%%%%%%%%%%%%%%%  
Since the definition of corner bins is arbitrary, we could relabel and get the same result; symmetry requires that $<p_{1}^q>_{\alpha}=<p_{2}^q>_{\alpha}$.  These results can now be assembled to calculate the 1D scaled entropy for scales larger than the support size
%%%%%%%%%%%%%%%%%%%%%%%%%%%%%%%%%%%%%%%%%%%%%%%%%%%%%%%%%%%%%%%
\begin{eqnarray}
S_q(e)&=&\frac{1}{1-q} \log [C_q(e)] \\ \nonumber
&=&\frac{1}{1-q} \log \left[1-\frac{L}{e}+\frac{L}{e}\left(\frac{2}{1+q}\right)\right] \\ \nonumber
&=&\frac{1}{1-q} \log \left[1+\left(\frac{1-q}{1+q}\right)\frac{L}{e}\right].
\end{eqnarray}
%%%%%%%%%%%%%%%%%%%%%%%%%%%%%%%%%%%%%%%%%%%%%%%%%%%%%%%%%%%%%%%

Combining results, we obtain the rank-$q$ scaled entropy for each degree of freedom of a BUD over all scales
%%%%%%%%%%%%%%%%%%%%%%%%%%%%%%%%%%%%%%%%%%%%%%%%%%%%%%%%%%%%%%%
\begin{equation} \label{contin-ent}
S_q(e)=\cases{\log\left(\frac{L}{e}\right)+\frac{1}{1-q}\log\left[1+\left(\frac{1-q}{1+q}\right)\frac{e}{L}\right],&for $e \le L$\cr
\frac{1}{1-q} \log\left[1+\left(\frac{1-q}{1+q}\right)\frac{L}{e}\right],&for $e \ge L$.\cr}		
\end{equation}
%%%%%%%%%%%%%%%%%%%%%%%%%%%%%%%%%%%%%%%%%%%%%%%%%%%%%%%%%%%%%%%
We have generated the exact expression for the scale-local entropy in the general case of a $d$-dimensional bounded uniform distribution.

\section{Point Sets as BUDs}

Eq. (\ref{contin-ent}) was derived in the continuum limit, it does not describe a uniform random point set. We generalize to the reference for discrete distributions by introducing an additional factor in Eq. (\ref{contin-ent}).  The entropy of a bounded random uniform distribution (BRUD) follows the BUD reference at larger scales but eventually approaches $\log (N)$ as a limiting value for scales smaller than the typical two-point separation.  The BUD reference can be generalized to the discrete case if we account for the appearance of void bins at smaller scales.  The fraction of void bins at scale $e$ for a Poisson-distributed point set is $\exp [-\mu_0(e)]$, where $\mu_0(e)\equiv Ne/L$ is the average bin occupancy for all bins within the boundary. The fraction of {\em occupied} bins (the support of the point set) is then $f(e)=1-\exp (-Ne/L)$. Incorporating $f(e)$ into the BUD entropy yields the BRUD result
%%%%%%%%%%%%%%%%%%%%%%%%%%%%%%%%%%%%%%%%%%%%%%%%%%%%%%%%%%%%%%%
\begin{equation} \label{point-ent}
S_q(e)=\cases{\log\left(\frac{L}{e}\right)+\frac{1}{1-q}\log\left[1+\left(\frac{1-q}{1+q}\right)\frac{e}{L}\right] + \log\left[1-\exp\left(\frac{-Ne}{L}\right)\right], ~~~ e \le L\cr
\frac{1}{1-q} \log\left[1+\left(\frac{1-q}{1+q}\right)\frac{L}{e}\right] + \log\left[1-\exp\left(\frac{-Ne}{L}\right)\right], ~~~ e \ge L.\cr}		
\end{equation}
%%%%%%%%%%%%%%%%%%%%%%%%%%%%%%%%%%%%%%%%%%%%%%%%%%%%%%%%%%%%%%%

%%%%%%%%%%%%%%%%%%%%%%%%%%%%%%%%%%%%%%%%%%%%%%%%%%%%%%%%%%%
%  3
% 50k RUD with BUD reference
% 
\begin{figure}[ht]
%\centereps{4in}{2in}{fig3new.eps}
\centering
\includegraphics[width=4in]{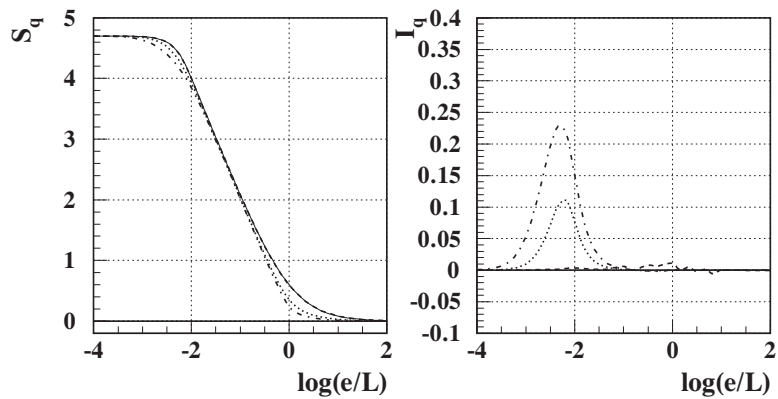}
\caption{Scale-local entropy and information for a 2D uniform random distribution of 50k points ($q=0$ dashed, $q=2$ dotted, $q=5$ dot-dashed lines) with analytic BRUD entropy for $q=0$ (solid line) as a reference.  Data and reference entropy distributions differ near the mean interparticle spacing scale ($log(\sqrt{N^{-1}})\sim$-2.35) as discussed in the text.}
\label{50kRUD}
\end{figure}
%
%%%%%%%%%%%%%%%%%%%%%%%%%%%%%%%%%%%%%%%%%%%%%%%%%%%%%%%%%%%

This entropy is precise for a discrete, random point set and $q=0$.  For $q > 0$ the reference entropy differs from the entropy for real point sets in a small scale interval near $e \approx L / \sqrt{N}$, the typical two-point separation. This information (entropy difference) derives from the fact that the form of R\'enyi entropy employed here is based on ordinary moments (averaged powers of $p_i(e)$), whereas the point-set data are Poisson distributed.  A correlation integral based on factorial moments would give zero information for uniform (uncorrelated) point sets \cite{SLTM}.

We could conclude that a factorial-moment approximation to the correlation integral should be used for point sets at least, if not other applications. However, extending our previous observation that boundedness itself is a form of correlation in a self-consistent system we can also view the point set as a result of increasing correlation (coalescence) of a continuous distribution at a characteristic scale. The point set does have correlations additional to the boundedness of the continuous BUD. The apparent discrepancies in the entropies of BRUD and real points sets (the nonzero $I_q$ in Fig. \ref{50kRUD}) reveal a genuine correlation feature in a discrete point set compared to a continuum. The R\'enyi entropies based on ordinary moments are preferred as a {\em more general} formulation applicable to arbitrary measure distributions. The information corresponding to the difference between factorial moments and ordinary moments for the uniform random point set is meaningful and can be expressed analytically (the subject of a future paper).

\section{Scale-local Dimension}

Having obtained entropy and information as scale-local distributions we can similarly express the dimension of a distribution as a function of scale.  We start with a conventional definition of dimension based on asymptotic limits \cite{Gollub}
%%%%%%%%%%%%%%%%%%%%%%%%%%%%%%%%%%%%%%%%%%%%%%%%%%%%%%%%%%%%%%%
\begin{equation}
d_q \equiv \frac{1}{q-1}\lim_{e\rightarrow 0} \Biggl[ \frac{\log \left(\sum_{i=1}^{M(e)}p_i^q(e)\right)}{\log (e)} \Biggr] = \lim_{e\rightarrow 0} \Biggl[ -\frac{S_q(e)}{\log (e)} \Biggr].
\end{equation}
%%%%%%%%%%%%%%%%%%%%%%%%%%%%%%%%%%%%%%%%%%%%%%%%%%%%%%%%%%%%%%%
In this approach dimension is a single number defined in the limit of a zero-scale partition. The slope of the $S_q(e)$ distribution is evaluated in the limit of zero scale. This definition favors certain specific types of correlation (power laws), and assumes that the limit exists, which may not be true even in principle. For more general cases the results can be misleading.  We relax the asymptotic limit restriction as we did with scale-local entropy to obtain a more general {\em scale-local} dimension
%%%%%%%%%%%%%%%%%%%%%%%%%%%%%%%%%%%%%%%%%%%%%%%%%%%%%%%%%%%%%%%
\begin{eqnarray}
d_q(e)&=&\lim_{\delta e\rightarrow 0} \Biggl[ -\frac{S_q(e+\delta e) -S_q(e) }{\log (e+\delta e) - \log(e)} \Biggr] = -\frac{\partial [S_q(e)]}{\partial [\log (e)]}.
\end{eqnarray}
%%%%%%%%%%%%%%%%%%%%%%%%%%%%%%%%%%%%%%%%%%%%%%%%%%%%%%%%%%%%%%%
Applying this definition to the generalized entropy of a BRUD in Eq. (\ref{point-ent}) yields
%%%%%%%%%%%%%%%%%%%%%%%%%%%%%%%%%%%%%%%%%%%%%%%%%%%%%%%%%%%%%%%
\begin{equation}
d_q(e)=\cases{1 - \frac{1}{1-q} \cdot \frac{\left(\frac{1-q}{1+q}\right)\frac{e}{L}}{1+\left(\frac{1-q}{1+q}\right)\frac{e}{L}} - \frac{\frac{Ne}{L} ~ \exp (-\frac{Ne}{L})}{1-\exp (-\frac{Ne}{L})}, ~~~ e \le L\cr
\frac{1}{1-q} \cdot \frac{\left(\frac{1-q}{1+q}\right)\frac{L}{e}}{1+\left(\frac{1-q}{1+q}\right)\frac{L}{e}} - \frac{\frac{Ne}{L} ~ \exp (-\frac{Ne}{L})}{1-\exp (-\frac{Ne}{L})}, ~~~ e \ge L.\cr}		
\end{equation}
%%%%%%%%%%%%%%%%%%%%%%%%%%%%%%%%%%%%%%%%%%%%%%%%%%%%%%%%%%%%%%%

%%%%%%%%%%%%%%%%%%%%%%%%%%%%%%%%%%%%%%%%%%%%%%%%%%%%%%%%%%%
%  4
% 2d scaled dimension example
% 
\begin{figure}[ht]
%\centereps{4in}{2in}{2d50kd.eps}
\centering
\includegraphics[width=4in]{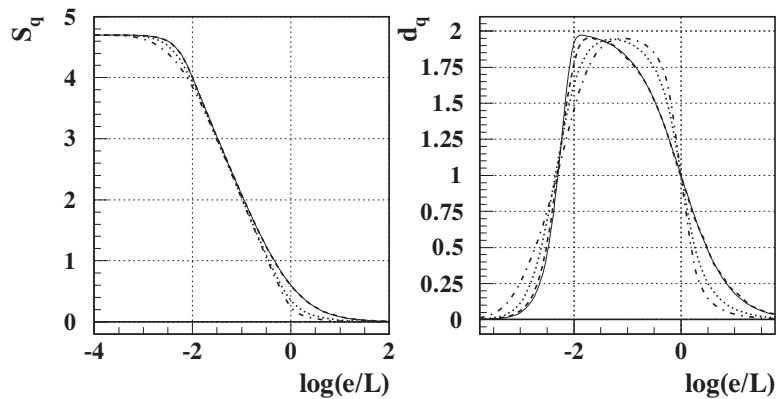}
\caption{Entropy and dimension plotted as a function of partition scale for a randomly generated uniform distribution ($q=0$ dashed, $q=2$ dotted, $q=5$ dot-dashed lines).  The solid lines shows the analytical $q=0$ results derived for a generalized BRUD.}
\label{Dq}
\end{figure}
%
%%%%%%%%%%%%%%%%%%%%%%%%%%%%%%%%%%%%%%%%%%%%%%%%%%%%%%%%%%%

Dimension expressed as a scale-local distribution is a novel aspect of this entropy treatment.  To understand how scale can affect the inferred dimension of an object consider the apparent dimension of a planet in the solar system from different viewpoints.  For an observer on one planet other planets appear to the eye as isolated points (with zero dimension).  With a powerful telescope the resolution size (scale) of the observation decreases substantially. Planets appear as 2D disks with 1D border.    A radar probe orbiting a planet can determine that the planet at smaller scale has a rich 3D surface and internal structure that could not be supported by a 1D point or 2D plane. Continuing to the atomic scale planets are made of atoms and molecules that appear point-like.  At this scale a planet's dimensionality returns to zero.  This general principle is illustrated in figure~\ref{Dq}.

\section{Analysis of Hierarchically Organized Point Sets}

As an exercise in precision correlation analysis using scale-local entropy and dimension we model cluster formation via condensation by generating a hierarchical point distribution with correlation features distributed over a range of scales.  This model is relevant to phase transitions and complex systems analysis. To create a two-dimensional, two-tier cluster hierarchy on a square region of side $L$ we generate a uniform random distribution of $N_0$ cluster sites, providing correlations at the characteristic length scale $L/\sqrt{N_0}$ -- the mean site separation.  At each cluster site we throw a randomly generated uniform distribution of $N_1$ points with width $\delta_1$, giving the distribution a second characteristic length scale $\delta_1/\sqrt{N_1}$.  

%%%%%%%%%%%%%%%%%%%%%%%%%%%%%%%%%%%%%%%%%%%%%%%%%%%%%%%%%%%
%  5
% 2d 2-step hierarchy
% 
\begin{figure}[ht]
%\centereps{4in}{4in}{h1010new.eps}
\centering
\includegraphics[width=4in]{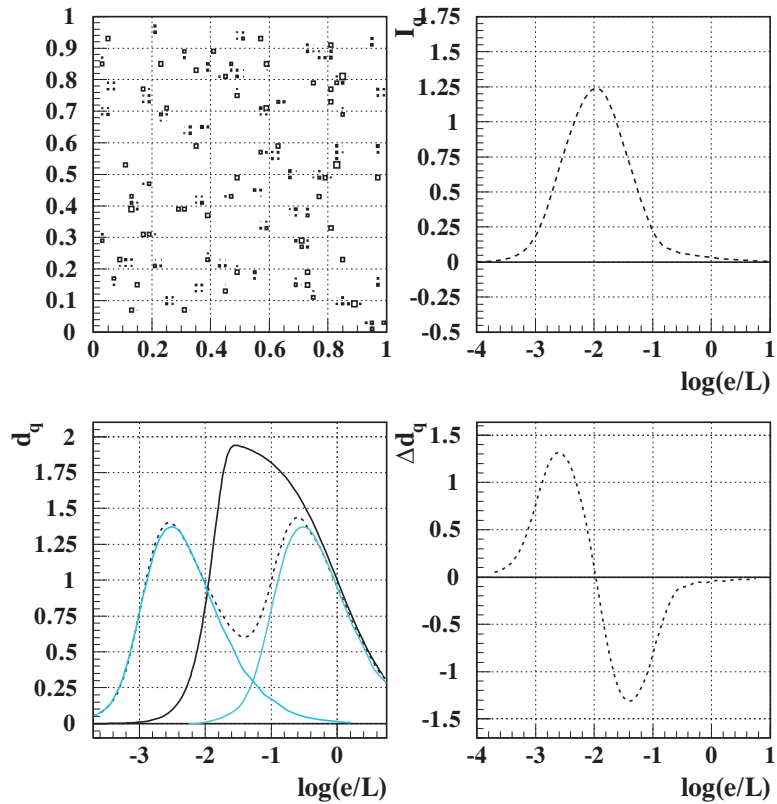}
\caption[Analysis results for a two-tier hierarchy.]{Analysis results for a two-tier hierarchy with $N_0=100$, $N_1=100$ \& $\delta_1=0.001$.  The scaled dimension of the data (dashed line) is compared to the analytic BRUD reference with the corresponding number of points ($N=N_0 N_1=10,000$) at scale $e/L=1$ (black solid line) as well as the analytic reference for $N=N_0=N_1=100$ at scales $1$ and $0.001$ (gray solid lines).}
\label{2d2stepH}
\end{figure}
%
%%%%%%%%%%%%%%%%%%%%%%%%%%%%%%%%%%%%%%%%%%%%%%%%%%%%%%%%%%%

If the two tiers of the hierarchy are sufficiently separated on scale, as in Fig. \ref{2d2stepH}, the sub-structure of the clusters is not evident to the analysis at large scale ($e \gg \delta_1$) .  The distribution appears to be a BRUD of $N_0$ random points.  At smaller scale ($e \sim \delta_1$) the apparent structure is dominated instead by the internal cluster structure, a BRUD of $N_1$ points.  This two-tiered hierarchical distribution is a first approximation to a self-similar distribution: it appears as a BRUD at scales $L$ and $\delta_1$ simultaneously (see Fig.~\ref{2d2stepH}).  Extending the hierarchy by recursive self-similar cluster generation would converge to the limiting case of a fractal point distribution over an arbitrarily large scale interval ({\em e.g.,} Cantor set). The `fractal' dimension would depend on the relations among the hierarchy scale separation, the cluster size and the point count.

The lower right panel of figure~\ref{2d2stepH} shows the {\em dimension transport} for the two-tier hierarchy.  Dimension transport is defined as the scale derivative of information, $\Delta d_q(e) \equiv -\partial [I_q(e)]/\partial [\log (e)]$, and is a measure of the scale-dependent dimension difference between reference and object distributions. Dimension transport measures increasing correlation as the transport of dimensionality from larger to smaller scale.  In the example of the two-tier hierarchy correlation (relative to the BRUD of the same multiplicity) anticorrelation of points at larger scale is achieved when points condense toward the cluster sites.  The anti-correlation at larger scale results in a reduction of larger-scale dimensionality (the system appears more point-like) and an increase in the local point density and dimensionality at smaller scale.  

%%%%%%%%%%%%%%%%%%%%%%%%%%%%%%%%%%%%%%%%%%%%%%%%%%%%%%%%%%%
%  6
% 6 panel hierarchy results
% 
\begin{figure}[ht]
%\centereps{4in}{4.412in}{c3f6.eps}
\centering
\includegraphics[width=4in]{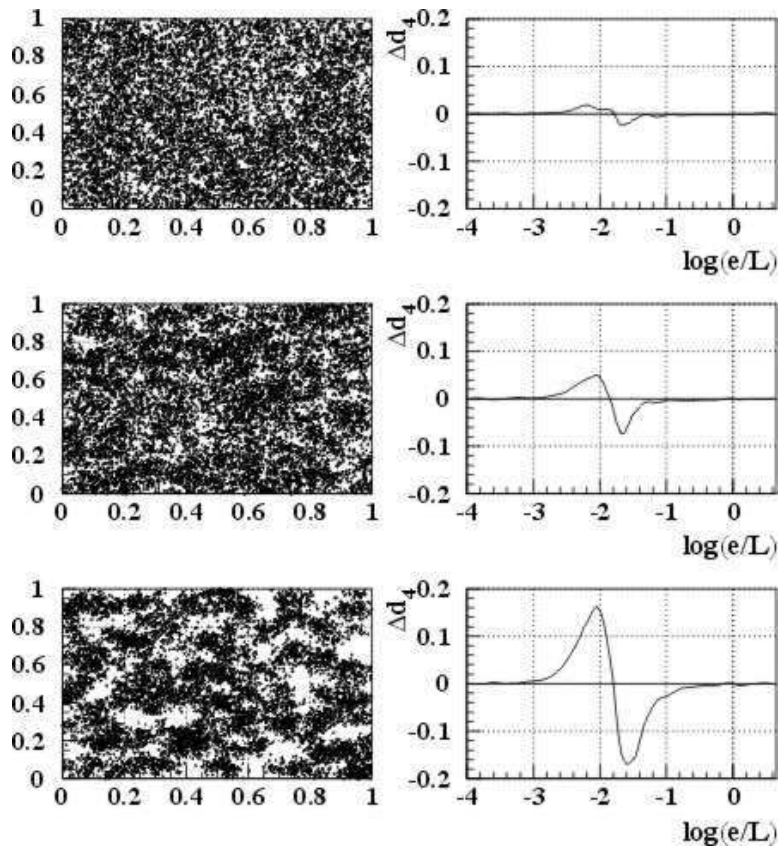}
\caption[Gradual onset of cluster formation in a system with a fixed total point multiplicity.]{A picture of the gradual onset of cluster formation ($\delta_1=0.03$) in a system with a fixed multiplicity ($N=10,000$).  Data are compared to a reference BRUD with equal multiplicity.}
\label{6panelH}
\end{figure}
%
%%%%%%%%%%%%%%%%%%%%%%%%%%%%%%%%%%%%%%%%%%%%%%%%%%%%%%%%%%%

The extended two-tier condensation example in figure~\ref{6panelH} shows how small-scale correlations increase by condensing points of a BRUD onto cluster sites.  At the onset of cluster formation ($\sim$3000 cluster sites, $\sim$3 points per cluster) the transport of dimension to smaller scale is barely visible (but still non-statistical).  When the size of the clusters becomes significant ($\sim$1000 cluster sites, $\sim$10 points per cluster) the analysis indicates what the eye perceives directly, that the distribution of points is in some way correlated.  When the cluster size is 10\% of the number of clusters ($\sim$300 cluster sites, $\sim$30 points per cluster) the dimension transport shows quite dramatically and quantitatively the transport of dimension from larger to smaller scale.

\section{Summary and Conclusions}

We have developed a novel analysis system which is well-suited to the task of precision correlation analysis for general measure distributions and especially for systems which exhibit clustering or other self-similar behavior.  By extending the R\'enyi-entropy concept to a locally-defined function of scale we are able to establish a more complete picture of data correlations and make precision comparisons among data, simulations and model distributions in the context of information theory.  Comparison of Monte Carlo results and analytic distributions have led to a detailed understanding of scale-local entropy measures.  Analysis of simulated clustering data suggests the power of this method in the quantification of scale-dependent correlation structure.

\section{Acknowledgments}

The authors would like to thank all of the people who have contributed to and supported this work.  In particular, we would like to thank Dhammika Weerasundara who was instrumental in the development of scale-local entropy methods, as well as Stephen Bailey, Justin Prosser, and Curtis Reynolds who helped implement several applications of this analysis.

%% :::::::::::::::::::::::::::::::::::::::::::::::::::::::::::::::::::::::


\section{References}

\begin{thebibliography}{5}

\bibitem{Gollub} Baker G L and Gollub J P 1990 {\em Chaotic Dynamics} (Cambridge: Cambridge University)
\bibitem{Grassberger} Grassberger P 1983 {\em Phys. Lett.\/} A {\bf 97} 224-230
\bibitem{KandK} Kittel C and Kroemer H 1980 {\em Thermal Physics} (New York: Freeman)
\bibitem{Wavelet} Pando J and Fang L 1998 {\em Phys. Rev.} E {\bf 57} 3553-3601
\bibitem{Pantin} Pantin E and Starck J-L 1996 {\em Astron. Astrophys. Suppl. Ser.} {\bf 118} 575-585
\bibitem{Thesis} Reid J G 2002 {Event-by-event Analysis Methods and Applications to Relativistic Heavy-ion Collision Data} {\em Doctoral Dissertation} nucl-ex/0302001
\bibitem{Renyi} R\'enyi A 1960 {\em MTA III. Oszt. Koezl.} {\bf 10} 251-282
\bibitem{SLTM} Trainor T A 1998 {Scale-local Topological Measures}, {\em Preprint}, University of Washington, 1998

\end{thebibliography}
\end{document}